\documentclass{aa} 
\usepackage{epsf}
\usepackage{graphicx}

\newcommand{\mmin}{$M_{\rm min}$}
\newcommand{\msun}{$M_{\odot}$}

\newcommand{\teff}{$T_{\rm eff}$}
\newcommand{\erg}{ergs\,s$^{-1}$}
\newcommand{\ergs}{ergs\,cm$^{-2}$s$^{-1}$}
\newcommand{\ergsa}{ergs\,cm$^{-2}$s$^{-1}$\AA$^{-1}$}

\newcommand{\mk}{$M_{\rm K}$}

\newcommand{\flam}{$f_{\lambda}$}
\newcommand{\Flam}{$F_{\lambda}$}
\newcommand{\ftio}{$f_{\rm TiO}$}
\newcommand{\Ftio}{$F_{\rm TiO}$}

\newcommand{\tio}{$\lambda 7500/7165$\AA{}}
\newcommand{\porb}{$P_{\rm orb}$}
\newcommand{\pmin}{$P_{\rm min}$}
\newcommand{\ha}{H$\alpha$}

\newcommand{\smyr}{$M_{\odot}$yr$^{-1}$}
\begin{document}
\thesaurus{08.05.3, 08.06.2, 08.12.2, 8.14.2, 8.23.1}
\title{Evidence for a substellar secondary in the magnetic cataclysmic
binary EF Eridani \thanks{Based in part on observations at the
European Southern Observatory La Silla (Chile)} }
\author{K. Beuermann \inst{1} 
\and P. Wheatley \inst{2}
\and G. Ramsay \inst{3} 
\and F. Euchner \inst{1} 
\and B.T. G\"ansicke \inst{1}} 
\offprints{beuermann@uni-sw.gwdg.de}
\institute{ 
Universit\"ats-Sternwarte, Geismarlandstr. 11, D-37083 G\"ottingen, Germany
\and X-ray Astronomy Group, Dept. of Physics and Astronomy, Leicester
University, University Road, Leicester LE1\,7RH, UK
\and Mullard Space Science Laboratory, University College of London,
Holmbury St. Mary, Dorking, Surrey RH5\,6NT, UK}
\date{Received December 6, 1999 / Accepted December 21, 1999}
\authorrunning{K. Beuermann et al.}
\titlerunning{Substellar secondary in EF Eri}
\maketitle
\begin{abstract}
Low-state spectrophotometry of the short-period polar EF Eridani
(\porb = 81 min) found the system at $V = 18.0$ with no trace of the
companion (Wheatley and Ramsay 1998). We show that the lack of such
spectral features
implies that the companion to the white dwarf in EF Eri has a spectral
type later than M9 and is either a transition object at the brink of
hydrogen burning or a brown dwarf. The optical low state spectrum
indicates a temperature of the white dwarf of
\teff$=9500\pm500$\,K. This is one of the coldest white dwarfs in
cataclysmic variables, implying a cooling age $t_{\rm
cool}\,\ga\,10^9$\,yrs or accretional heating at a rate as given by
gravitational radiation.  The large age of the system excludes a warm
brown dwarf as companion. EF Eri has either just passed through the
period minimum of cataclysmic variable stars or has started mass
transfer from an old brown dwarf secondary.

\end{abstract} 
\section{Introduction}
 
Cataclysmic binary stars (CVs) with short orbital periods
(\porb$\,<\,2$\,h) lose angular momentum by gravitational radiation,
causing these systems to shrink and their orbital periods to
decrease. Mass loss ablates the secondary star until it drops below
the minimum mass needed for hydrogen burning, becomes increasingly
degenerate, and enters the regime of brown dwarfs. The associated
change in the mass--radius relation of the secondary causes the
orbital period \porb\ to increase again. The abrupt cutoff in the
period distribution of CVs locates the observed period minimum at
77\,min (e.g. Kolb \&\ Baraffe 1999).
Short-period CVs with \porb$\,>\,90$\,min which approach the period 
minimum have observed secondaries with spectral types dM4--dM6. For
\porb$\,<\,90$\,min, the secondary stars are expected to be of still
later spectral type and have escaped detection so far (e.g. Beuermann
et al. 1998). Some dwarf novae of the WZ Sge type 
are suspected to have passed the period minimum and to harbour a
degenerate companion (Howell et al. 1995).  V592 Her may be such a
system (van Teeseling et al. 1999).

Direct observational proof of this evolutionary scenario is scarce so
far.  Spectroscopic detection of the secondaries in short-period dwarf
novae is extremely difficult because the accretion disc and the white
dwarf dominate the optical/near IR emission of the systems. The
situation is more favourable in magnetic CVs which lack a disc. In
these systems, accretion ceases at irregular instances and reveals the
naked binary.

In this paper, we discuss low state observations of the 81-min 
magnetic CV EF Eri (Wheatley \& Ramsay 1998) which allow us to set 
tight limits on the spectral flux and on 
the physical nature of the secondary.

\section{Observations}

\begin{figure*}[t]
\includegraphics[width=12cm]{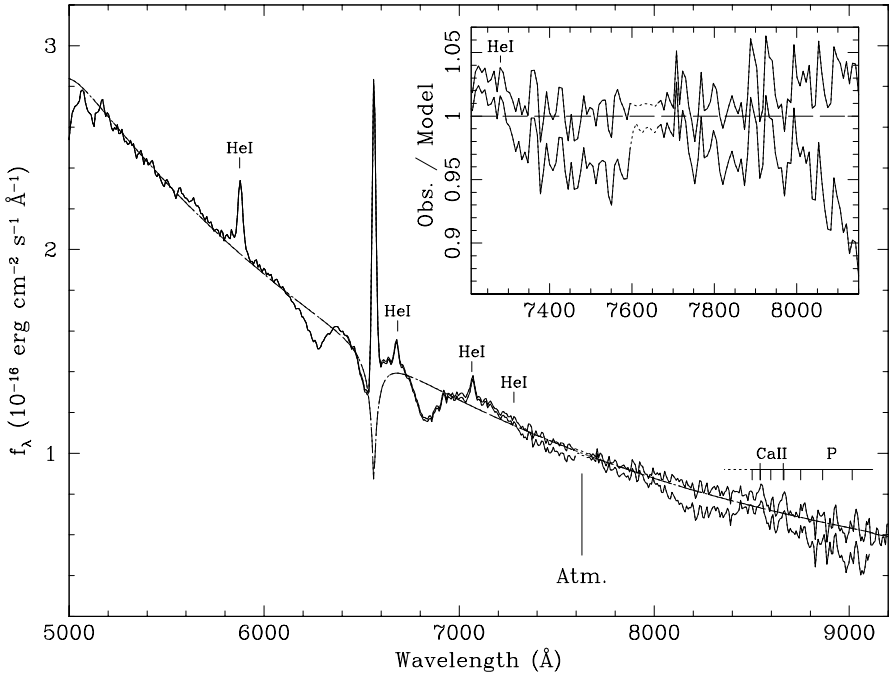}
\hfill
\raisebox{6mm}{
\begin{minipage}[b]{52mm}
\caption[]{\label{mbol_radius} Mean spectrum of EF Eri in the 1997 low
state for orbital phases corrected for the remnant cyclotron emission
(upper solid curve) along with the best-fit non-magnetic model
spectrum for \teff\,=\,9500\,K (dot-dashed curve). The lower solid curve
indicates the observed spectrum corrected for a 5\% flux contribution
at 7500\AA\ by a dM9 companion star. The negative M-star features indicate
that the companion must be fainter than this. The spectral region
between 7600 and 7660\AA\ is uncertain due to the correction for the
atmospheric \mbox{A-band} (dashed section of the curves).}
\end{minipage}}
\end{figure*}

Wheatley \&\ Ramsay (1998) found EF Eri in a low state in February
1997 with V varying between 17.95 and 18.20. Besides photometry, they
performed spectrophotometry on February 7 and 8 using the ESO 3.6-m
telescope at La Silla/Chile equipped with the focal reducer
spectrograph EFOSC\,1 Accretion did not cease completely as is
evidenced by Balmer line emission and weak cyclotron emission in the
red/infrared part of the spectrum. Cyclotron emission was observed
only during linear polarization phases 0.5--0.9 and was removed from
the individual spectra using the difference spectrum between cyclotron
maximum and the cyclotron free spectra as a template.  In the two
nights, the observations started at airmasses 1.1 and 1.2 and were
carried on until airmasses 1.8 and 2.5, respectively. A slow flux
decrease was observed with increasing airmass, probably caused by the
deteriorating seeing at lower altitude.  We adopted the flux level of
the cyclotron-free spectra at the beginning of the observations
(airmass $<1.27$) and corrected this level to zero airmass.
A nearby standard star of early spectral type was used to correct for
the telluric O$_2$ and H$_2$O bands. This correction minimizes the
excursions from a smooth continuum and causes the resulting spectrum
to be artificially smooth at the centre of the \mbox{A-band} between
7600 and 7660\AA.  The resulting mean spectrum is shown in Fig.\,1 as
the upper solid line and the narrow uncertain section between 7600 and
7660\AA\ is shown as the dashed line. The pronounced dips at 6300,
6500, and 6800\,\AA\ are the $\sigma^+$, $\pi$, and $\sigma^-$ Zeeman
components of \ha.  Zeeman tomography of the magnetic-field structure
of EF Eri will be presented elsewhere.

\section{Spectral analysis}

We fit the continuum of the white dwarf at wavelengths $\lambda >
5300$\AA\ (excluding the \ha\ absorption and emission line region)
with pure-hydrogen line-blanketed magnetic and non-magnetic model
atmosphere spectra (Jordan 1992, G\"ansicke et al. 1995). The best fit
is achieved for a mean effective temperature of $9500\pm500$ K
(dashed curve in Fig.\,1).  The observed spectral flux and the surface
flux of the white dwarf for \mbox{\teff\,=\,9500\,K} gives an
angular radius \mbox{$R_1/d\,=\,(2.30\pm 0.17)\times 10^{-12}$}
radians. Using magnetic model atmospheres yields a temperature lower
by $200$\,K and an identical angular radius.  As in other polars, the
remnant modulation of the blue continuum of 0.2 mag (Wheatley \&\
Ramsay 1998) can be explained by a moderately hot spot of 15\,000\,K
covering 6\% of the cross section of the white dwarf.
%
%
The model spectrum excellently fits the observed continuum between
5300 and 9000\,\AA\ with no evidence for a contribution by the
companion.

We estimate the maximum possible contribution by the unseen companion,
using the calibration of the surface brightness of late type main
sequence stars in the prominent \tio\ TiO/VO band (Beuermann 1999, 
and paper in preparation). This calibration is based on the ZAMS
stellar model radii of Baraffe et al. (1998), parameterized by
Beuermann et al. (1999). Fig.\,2 shows the derived surface fluxes
$F_{\lambda}$ of the late-type dwarfs Gl644C = VB8 (M7), Gl752B = VB10
(M8), and LHS2065 (M9).
We denote the difference between the surface fluxes at 7500\AA\ and
7165\AA\ as \Ftio\ and have calibrated this quantity as a function of
the spectral type for the range dM0--dM9. Gl 644C, Gl752B, and LHS2065
have \Ftio\,=\, $6.7\times10^4,~3.0\times10^4$, and $1.2 \times
10^4$\,\ergsa, respectively, very close to the mean \Ftio\ values of
dM7, dM8, and dM9 dwarfs. For the present purpose, we include the very
cool dwarf 2MASS1439+19 of spectral type L1 (Kirkpatrick et al. 1999)
for which we estimate \Ftio$\,\simeq\,0.4\times10^4$\,\ergsa. The observed 
flux of the companion which corresponds to \Ftio\ is denoted \ftio.

We subtract test fractions of the spectra of these stars from the
observed spectrum and judge the acceptable contribution by eye. The
lower solid curve in Fig.\,1 shows the effect of an assumed
contribution of LHS2065 which amounts to 5\% of the observed flux at
7500\AA. The resulting wavy structure is due to the subtracted M-star
and clearly indicates that the contribution by the companion must be
smaller than this. The insert shows the region around 7500 \AA\ in
more detail, now depicted as a flux ratio relative to the model
spectrum. The signature of the M-star is clearly visible in the
subtracted spectrum. We adopt the 5\% flux contribution at 7500 \AA\
as a strict upper limit to the flux contributed by the companion which
corresponds to upper limits on \ftio\ of 4.5, 3.9, 3.5 and
$2.3\times10^{-18}$\,\ergsa\ for spectral types M7, M8, M9, and L1,
respectively. 

\begin{figure}[t] 
\mbox{\includegraphics[width=8.8cm]{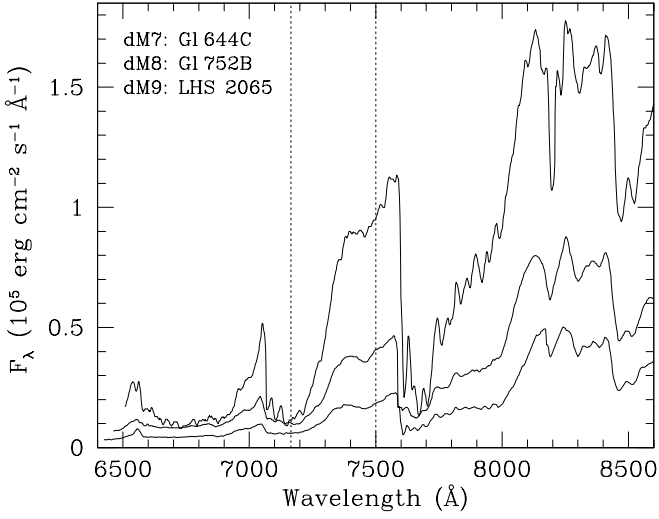}}
\caption[]{\label{mbol_radius}Surface fluxes \Flam\ of representative
late type dwarfs of spectral types dM7 (Gl644C), dM8 (Gl752B),
and dM9 (LHS\,2065). The vertical dashed lines indicate the
wavelengths at which the quantity \Ftio\ is measured (from Beuermann
1999).}
\end{figure}

\begin{table}[t]
\caption[ ]{Parameters of EF Eri for companions of spectral type dM7
to L1 and for a white dwarf with \teff\,=\,9500\,K and a mass between
0.4 and 1.0\,\msun. $M_2$ is from Eq. (4) and $\dot M_{\rm hi}$ is in
units of $10^{-11}$\msun\,yr$^{-1}$.  Acceptable $M_1$ -- spectral
type combinations are shown in boldface.  }
\begin{flushleft}
\begin{tabular*}{\hsize}{@{\extracolsep{\fill}}cccccccl}
\noalign{\smallskip} \hline \noalign{\smallskip}
$M_1$  & $R_1$    &$d$ &$\dot M_{\rm hi}$&Spectral&  $R_2$ & $M_2$ & Notes\\[0.3ex] 
\msun  &$10^9$\,cm& pc && Type   & $10^{9}$cm  & \msun & \\[0.3ex] 
\noalign{\smallskip} \hline \noalign{\smallskip}
0.4 & 1.17 & 165 &  25  & M7 & 4.5 & $<0.012$ & 1,2,5 \\
    &      &     &      & M8 & 6.3 & $<0.032$ & 1,2,5 \\
    &      &     &      & M9 & 9.3 & $<0.104$ & 1\\
    &      &     &      & L1 &13.2 & $<0.302$ & 1,3 \\
{\bf 0.6} &  0.91 & 128 &  7.7 & M7 & 3.5 & $<0.006$ & 2,5 \\
    &      &     &      & M8 & 4.9 & $<0.015$ & 2,5 \\
    &      &     &      & M9 & 7.2 & $<0.049$ & 2,5 \\
    &      &     &      & {\bf L1} & 10.2 & $<0.142$ & {\bf 3} \\
    &      &     &      & {\bf $\ga$L4} &&          & {\bf 4} \\
{\bf 0.8} &  0.73 & 103 &  2.9 & M7 & 2.8 & $<0.003$ & 2,5 \\
    &      &     &      & M8 & 3.9 & $<0.008$ & 2,5 \\
    &      &     &      & M9 & 5.8 & $<0.025$ & 2,5 \\
    &      &     &      & L1 & 8.2 & $<0.073$ & 2,3,5 \\
    &      &     &      & {\bf $\ga$L4} &&          & {\bf 4} \\
{\bf 1.0} &  0.57 &  80 &  1.1 & M7 & 2.2 & $<0.002$ & 2,5 \\
    &      &     &      & M8 & 3.1 & $<0.004$ & 2,5 \\
    &      &     &      & M9 & 4.5 & $<0.012$ & 2,5 \\
    &      &     &      & L1 & 6.4 & $<0.035$ & 2,5\\
    &      &     &      & {\bf $\ga$L4} &&          & {\bf 4} \\
\noalign{\smallskip}\hline\noalign{\smallskip}
\end{tabular*}
{\bf Notes: } (1) $M_1 = 0.4$\,\msun\ excluded by high temperature of
hard X-ray spectrum; (2) brown dwarf with $M_2$ and assumed spectral
type exclu- ded at age ~$t\,>\,t_{\rm cool}$ (see text) (Baraffe et
al. 1998); (3) spectral type indicates a transition object with
$M_2\,\sim\,0.075$\,\msun; (4) spectral type implies a brown dwarf
companion; (5) $M_2$ may not be able to supply $\dot M_{\rm hi}$ if driven by
gravitational radiation only (Kolb \& Baraffe 1999).
\end{flushleft}
\end{table} 

%
%
%
%

\section{Companion mass}

We now derive mass limits for the companion in EF Eri as functions of
its spectral type and an assumed white dwarf mass $M_1$. Its radius
$R_1$ and the distance $d$ to EF Eri are related by
\begin{equation}
R_1/d = (f_{\lambda}/F_{\lambda})_{\rm wd}^{1/2}
\end{equation}
\noindent where \flam\ and \Flam\ are the observed continuum flux and
the model surface flux of the white dwarf with \teff\,=\,9500\,K,
respectively, and $R_1/d$ is as quoted above. Correspondingly, the
ratio of the (upper limit to the) observed flux \ftio\ and the surface
flux \Ftio\ of the companion in the \tio\ band is given by
\begin{equation}
R_2/d = (f_{\rm TiO}/F_{\rm TiO})_{\rm sec}^{1/2}.
\end{equation}
Finally, Roche geometry implies
\begin{equation} 
R_2 = 1.628\times10^{10}\,(M_2/M_{\odot})^{1/3}\,P_{\rm h}^{2/3}f(q) \qquad
{\rm cm}
\end{equation}  
\noindent where $P_{\rm h}$ is the orbital period in hours, $f(q)$ is a
function of the mass ratio $q = M_1/M_2$ which varies between 0.98 and
1.01 for $q$ between 3 and 30, and $M_1$ and $M_2$ are the masses of
the white dwarf and the companion, respectively.  Eqs. (1)\,--\,(3)
define (an upper limit to) the companion mass $M_2$ as a function of
the unknown mass of the white dwarf and the spectral type of the
companion, represented by $R_1$ and \Ftio, respectively,
\begin{equation} 
M_2 = \left(\frac{R_1}{f(q)\,1.628\times10^{10} {\rm cm}}\right)^3
\left(\frac{d}{R_1}\right)_{\rm wd}^3 \left(\frac{f_{\rm TiO}}{F_{\rm
TiO}}\right)_{\rm sec}^{3/2} \frac{1}{P_{\rm h}^2}.
\end{equation} 
\noindent 
We use radii of white dwarfs with \teff\,=\,9500\,K and an
envelope consisting of $10^{-4}$\msun\ of hydrogen and $10^{-2}$\msun\
of helium (D. Koester, private communication). Carbon white dwarfs of
1.0, 0.8, and 0.6 \msun\ have $R_1$ = 0.57, 0.73, and 0.91$\times
10^9$\,cm, respectively. An 0.4\,\msun\ helium white dwarf has $R_1 =
1.17\times 10^9$\,cm.  In order to obtain upper limits to $M_2$, we
employ the 1-$\sigma$ upper limit to $d/R_1$ as given above and the
upper limits to \ftio. Table 1 lists the resulting strict upper limits
to $M_2$ for four values of $M_1$ and the assumed spectral types of
the companion. Some insight into the \mbox{systematics} of the
\mbox{$M_2$-variation} may be gained from noting that for low-mass
white dwarfs with $R_1 \propto M_1^{-1/3}$, one obtains $M_2 \propto
M_1^{-1}F_{\rm TiO}^{-3/2}$, i.e. the allowed mass of the companion is
lowest for a massive white dwarf and an early spectral type of the
companion. For brown dwarfs, $M_2$ is limited by the assumed spectral
type and the age rather than by Eq. (4).

\section{Accretion rate}

Before its recent low state, EF Eri was known as a roughly steady
X-ray source for two decades. The \mbox{integrated} high-state
orbital mean X-ray and cyclotron flux is $f_{\rm hi}\,= 3\times
10^{-10}$\,\ergs\ (Beuermann et al. 1987, 1991). The soft X-ray flux,
still quite uncertain in the EINSTEIN observations (Beuermann et
al. 1987), is more accurately defined in the ROSAT data (Beuermann et
al. 1991). The integrated luminosity is $L_{\rm hi} \simeq
2.6\times 10^{32}\,(d/100\,{\rm pc})^2$\,\erg, where we have accounted
for the different geometry factors of the individual radiation
components following Beuermann et al. (1987). The corresponding
distance-dependent high-state accretion rate
\begin{equation} 
\dot M_{\rm hi} = L_{\rm hi}\,R_{\rm wd}/(G M_{\rm wd})
\end{equation} 
is probably accurate to better than a factor of two (Table\,1). We
expect that $\dot M_{\rm hi} > \langle\dot M\rangle$, the
long-term mean mass transfer rate which is about
$3\times 10^{-11}$\,\smyr\ for short-period CVs with low-mass main
sequence donors (Kolb \&\ Baraffe 1999, their Fig. 1).  For $M_1 \ge
0.8$\,\msun, $\dot M_{\rm hi}$ and even more so $\langle\dot 
M\rangle$ fall below $3\times 10^{-11}$\,\smyr, suggesting a substellar 
companion. For lower $M_1$, the derived value of $\dot M_{\rm hi}$ permits a 
stellar or a substellar companion.

\section{Discussion and Conclusion}

Table 1 lists the limits on $M_2$ obtained from Eq. (4) for an assumed
white dwarf mass and spectral type of the companion. We have included
a line for brown-dwarf companions with spectral type $\ga$\,L4 for which
Eq. (4) is no longer applicable.

Several authors have argued in favour of a rather massive white dwarf
in EF Eri. Beuermann et al. (1987) find $M_1 > 0.56$\,\msun, Wu et
al. (1995) $M_1 \simeq 0.57$\,\msun, and Cropper et al. (1999)
$M_1\,\simeq \,0.80$\,\msun, all using the observed bremsstrahlung
temperature. The observed mass function (Mukai \&\ Charles 1985)
suggests $M_1 \ga 0.8$\,\msun. Hence, $M_1 \ga 0.6$\,\msun\
and the companion is substellar or later than M9 (Note 1 in Table 1).

Stars just above the limit for hydrogen burning, \mmin$\,\simeq
0.075$\,\msun, with solar composition and an age $>2$\,Gyr reach a
stable effective temperature of $\sim2000$\,K (Baraffe et al. 1998).
Their nominal spectral type would be $\sim$M9/L0, while objects of
slightly lower mass rapidly fade to still lower temperatures. The
details are still uncertain because the definition of the
\mbox{M-star/L-dwarf} transition is a spectroscopic one and the
conversion to mass is not straightforward (Baraffe et al. 1998,
Kirkpatrick et al. 1999, Reid et al. 1999).

Another important result of this study is the low effective
temperature of the white dwarf in EF Eri. With
\teff$\,\simeq\,9500$\,K it is the coldest of all white dwarfs in CVs
which have a reliably determined temperature (G\"ansicke 1999).
This low temperature implies a cooling age of $t_{\rm cool} =
9\times10^8-2\times10^9$\,yrs for $M_1 = 0.6-1.0$\,\msun, respectively
(Wood 1995). The companion must be older than $t_{\rm cool}$ and,
hence, can not be a warm brown dwarf.  This excludes all combinations
of a spectral type M7--L1 and $M_2 < 0.06$\,\msun\ (Note 2 in Table
1). For definiteness, we identify a spectral type L1 with a transition
object of mass $M_2 \sim 0.075$\,\msun\ and a spectral type $\ga$\,L4
with a brown dwarf (Notes 3 and 4 in Table 1).  Acceptable companion
objects are those which have only a boldface entry {\bf 4} in the
Notes column of Table 1 (brown dwarfs). Acceptable is also the sole
entry {\bf 3} (transition object) for $M_1 = 0.6$\,\msun, but
marginally so because a companion of $\sim$0.075\,\msun\ is possibly
not able to supply $\dot M_{\rm hi} = 8\times 10^{-11}$\,\smyr (Note 5). We
conclude that the companion is either a transition object or a brown
dwarf.

The true age of the white dwarf must exceed $t_{\rm cool}$ because the
temperatures of accreting white dwarfs are kept above those of field
white dwarfs of the same age by compressional heating of their
envelopes (Sion 1995, Warner 1995, G\"ansicke et al. 1999). If
entirely due to accretion, \teff\,=\,9500\,K requires a mean accretion
rate not exceeding that expected from gravitational radiation,
$\langle\dot M\rangle \la 3\times 10^{-11}$\,\ergs\ (G\"ansicke 1999),
lending further support to a rather high white dwarf mass and a
substellar companion.

The substellar nature of the companion is supported also by
considerations of the evolutionary state of EF Eri. One of the main
results of the theory of CV evolution is the recognition that a period
bounce occurs when $M_2 \simeq 0.06$\,\msun\ and the secondary
becomes increasingly degenerate. The orbital period of EF Eri is 4 min
longer than the observed minimum period \pmin$ \simeq 77$\,min and
still further above the theoretical value of 67\,min (Kolb \&\ Baraffe
1999). If EF Eri evolved from longer periods, it is now either
approaching the minimum or has already passed it. At 4 min above the
{\em calculated} period minimum the donor mass/temperature would be
either 0.08\,\msun/2500\,K or 0.05\,\msun/1500\,K, while at a
calculated period of 81\,min these numbers would be
0.10\,\msun/2900\,K and 0.035\,\msun/1000\,K (Kolb \&\ Baraffe
1999). Since a spectral type L
implies \teff$ \la 2000$\,K, we conclude that EF Eri has passed
\pmin. If it formed with a brown dwarf secondary of age $>
1$\,Gyr, the companion would have \teff$ \la 2000$\,K at
\porb\,=\,81\,min, too (Kolb \& Baraffe 1999). 
In both cases the companion is substellar with $M_2 \la 0.06$\,\msun.

The reason for the discrepancy between calculated and observed \pmin\
is not yet understood. Possible explanations include (i) the presence
of an additional mechanism of angular momentum loss and (ii) a
mechanism which causes CVs with stellar secondaries to become
unobservable before they reach \pmin. As the most likely value
of the companion mass is $<0.06$\msun\ and EF Eri is still a bright
X-ray source our result argues against the second explanation.

EF Eri is a key object for our understanding of CV evolution.
Obtaining a trigonometric parallax would allow to determine accurate
values of the mass, luminosity, and accretion rate of the white dwarf.
Infrared spectrophotometry will allow to detect the companion
(\mk$\,<\,12$ for companions as late as L8) against the veiling flux
of the white dwarf (\mk$~\simeq\,12.5$).

\acknowledgements{ We thank the referee Ulrich Kolb for helpful
comments which improved the presentation.  This work was supported in
part by DLR/BMBF grant 50\,OR\,9903\,6.}

\end{document}